# New Evaluation of the QED Running Coupling and of the Muonium Hyperfine Splitting

**Stephan Narison**
Laboratoire de Physique Mathématique, Université de Montpellier 2, Place Eugène Bataillon 34095 - Montpellier Cedex 05, France and Center for Academic Excellence on Cosmology and Particle Astrophysics (CosPA), Department of Physics, National Taiwan University, Taipei, Taiwan, 10617 Republic of China.
Email: narison@lpm.univ-montp2.fr

We present a new independent evaluation of the hadronic and QCD contributions to the QED running coupling $\alpha(M_Z)$ and to the muonium hyperfine splitting $\nu$. We obtain: $\Delta\alpha_{\text{had}} = 2770(17) \times 10^{-5}$ and $\Delta\nu_{\text{had}} = 232.5(2.5)$ Hz. Combined with the QED and Electroweak Standard Model contributions, they lead to: $\alpha^{-1}(M_Z) = 128.926(25)$ and to the Fermi energy splitting $\nu_F = 4\ 459\ 031\ 783(229)$ Hz , where for the latter, we have used, in addition, the precise measurement of the muonium hyperfine splitting $\nu_{\text{exp}}$. We use $\nu_F$ in order to predict the ratios of masses $m_\mu/m_e = 206.768\ 276(11)$ and of the magnetic moments $\mu_\mu/\mu_B^e = 4.841\ 970\ 47(25) \times 10^{-3}$, which are in excellent agreement with the ones quoted by the Particle Data Group. These remarkable agreements can provide strong constraints on some contributions beyond the Standard Model.

## 1 Introduction

A natural and important extension of our recent work [1] (hereafter referred as SN) on the hadronic and QCD contributions to the anomalous magnetic moment of the muon and tau leptons $a_l$, is the evaluation of the hadronic and QCD contributions to the QED running coupling $\alpha(M_Z)$ and to the muonium hyperfine splitting $\nu$. The analysis is important as it will give a complete set of estimates of three independent observables within the same inputs used in SN. These self-contained results should pass tests from a comparison with high precision experiments and different existing predictions in the literature. Using a dispersion relation, it is remarkable to notice that the different lowest order hadronic contributions for these three processes can be expressed in a closed form as a convolution of the $e^+e^- \to$ hadrons cross-section $\sigma_H(t)$ with a QED kernel function $K(t)$ which depends on each observable:

$$\mathcal{O}_{\text{had}} = \frac{1}{4\pi^3} \int_{4m_\pi^2}^{\infty} dt\ K_\mathcal{O}(t)\ \sigma_H(t)\ , \qquad (1)$$

where:

$$\mathcal{O}_{\text{had}} \equiv a_{l,\text{had}}\ , \quad \Delta\alpha_{\text{had}} \times 10^5 \quad \text{or} \quad \Delta\nu_{\text{had}}\ . \qquad (2)$$

− For the anomalous magnetic moment $a_{l,\text{had}}$, $K_{a_l}(t \geq 0)$ is the well-known kernel function [2]:

$$K_{a_l}(t) = \int_0^1 dx \frac{x^2(1-x)}{x^2 + (t/m_l^2)(1-x)}\ , \qquad (3)$$

where $m_l$ is the lepton mass. It behaves for large $t$ as:

$$K_{a_l}(t \gg m_l^2) \simeq \frac{m_l^2}{3t}\ . \qquad (4)$$

− For the QED running coupling $\Delta\alpha_{\text{had}} \times 10^5$, the kernel is (see e.g. [3]):

$$K_\alpha(t) = \left(\frac{\pi}{\alpha}\right)\left(\frac{M_Z^2}{M_Z^2 - t}\right)\ , \qquad (5)$$

where $\alpha^{-1}(0) = 137.036$ and $M_Z = 91.3$ GeV. It behaves for large $t$ like a constant.
− For the muonium hyperfine splitting $\Delta\nu_{\text{had}}$, the kernel function is (see e.g [4]):

where:
$$\rho_\nu = 2\nu_F \frac{m_e}{m_\mu}, \qquad x_\mu = \frac{t}{4m_\mu^2} \qquad v_\mu = \sqrt{1 - \frac{1}{x_\mu}}, \qquad (7)$$

and we take (for the moment) for a closed comparison with [4] [1], the value of the Fermi energy splitting:
$$\nu_F = 445\ 903\ 192\ 0.(511)(34)\ \text{Hz}. \qquad (8)$$

It behaves for large $t$ as:
$$K_\nu(t \gg m_\mu^2) \simeq \rho_\nu \left(\frac{m_\mu^2}{t}\right) \left(\frac{9}{2}\ln\frac{t}{m_\mu^2} + \frac{15}{4}\right). \qquad (9)$$

The different asymptotic behaviours of these kernel functions will influence on the relative weights of different regions contributions in the evaluation of the above integrals.

## 2 Input and Numerical Strategy

The different data input and QCD parametrizations of the cross-section $\sigma_H(t)$ have been discussed in details in SN [1] and quoted in the last column of Table 1, corresponding to the estimate in different regions. Table 1 is analogous to Table 2 of SN. We shall only sketched briefly the numerical strategy here:
− Our result from the $I = 1$ isovector channel below 3 GeV$^2$ is the mean value of the one using $\tau$-decay and $e^+e^-$ data. In both cases, we have used standard trapezoïdal rules and/or least square fits of the data in order to avoid theoretical model dependence parametrization of the pion form factor. In the region $0.6 - 0.8$ GeV$^2$ around the $\omega$-$\rho$ mixing, we use in both cases $e^+e^-$ data in order to take properly the $SU(2)_F$ mixing. The $SU(2)$ breaking in the remaining regions are taken into account by making the average of the two results from $\tau$-decay and $e^+e^-$ and by adding into the errors the distance between this mean central value with the one from each data.
− For the $I = 0$ isoscalar channel below 3 GeV$^2$, we use the contributions of the resonances $\omega(782)$ and $\phi(1020)$ using narrow width approximation (NWA). We add to these contributions, the sum of the exclusive channels from 0.66 to 1.93 GeV$^2$. Above 1.93 GeV$^2$, we include the contributions of the $\omega(1.42)$, $\omega(1.65)$ and $\phi(1.68)$ using a Breit-Wigner form of the resonances.
− For the heavy quarkonia, we include the contributions of known $J/\psi$ (1S to 4.415) and $\Upsilon$ (1S to 11.02) families and use a NWA. We have added the effect of the $\bar{t}t$ bound state using the leptonic width of $(12.5 \pm 1.5)$ keV given in [5].
− Away from thresholds, we use perturbative QCD plus negligible quark and gluon condensate contributions, which is expected to give a good parametrization of the cross-section. These different expressions are given in SN. However, as the relative rôle of the QCD continuum is important in the estimate of $\Delta\alpha_\text{had}$, we have added, to the usual Schwinger interpolating factor at order $\alpha_s$ for describing the heavy quark spectral function, the known $\alpha_s^2 m_Q^2/t$ corrections given in SN. However, in the region we are working, these corrections are tiny.
− On the $Z$-mass, the integral for $\Delta\alpha_\text{had}$ has a pole, such that this contribution has been separated in this case from the QCD continuum. Its value comes from the Cauchy principal value of the integral.

## 3 The QED Running Coupling $\alpha(M_Z)$

The result given in Table 1 corresponds to the lowest order vacuum polarization. Radiative corrections to this result can be taken by adding the effects of the radiative modes $\pi^0\gamma,\ \eta\gamma, \pi^+\pi^-\gamma, ....$ We estimate such effects to be:
$$\Delta\alpha_\text{had} = (6.4 \pm 2.7) \times 10^{-5} \qquad (10)$$

by taking the largest range spanned by the two estimates in [5] and [6]. Adding this (relatively small) number to the result in Table 1, gives the total hadronic contributions:
$$\Delta\alpha_\text{had} = 2769.8(16.7) \times 10^{-5}. \qquad (11)$$

Using the QED contribution to three-loops [3]:
$$\Delta\alpha_\text{QED} = 3149.7687 \times 10^{-5}, \qquad (12)$$

and the Renormalization Group Evolution of the QED coupling:
$$\alpha^{-1}(M_Z) = \alpha^{-1}(0)\Big[1 - \Delta\alpha_\text{QED} - \Delta\alpha_\text{had}\Big], \qquad (13)$$

---
[1]In the next section, we shall extract this value from the analysis.

Table 1: Lowest order determinations of $\Delta\alpha_{\rm had} \times 10^5$ and $\Delta\nu_{\rm had}$ [Hz] using combined $e^+e^-$ and inclusive $\tau$ decay data (2nd and 4th columns) and averaged $e^+e^-$ data (3rd and 5th columns).

| Region in GeV$^2$ | $\Delta\alpha_{\rm had} \times 10^5$ | | $\Delta\nu_{\rm had}$ [Hz] | | Data input |
|---|---|---|---|---|---|
| | $\tau+e^+e^-$ | $e^+e^-$ | $\tau+e^+e^-$ | $e^+e^-$ | |
| **Light Isovector** | | | | | |
| $4m_\pi^2 \to 0.8$ | $314.5 \pm 2.3$ | $302.7 \pm 7.1$ | $152.9 \pm 1.8$ | $148.4 \pm 3.1$ | [6, 7, 8] |
| $0.8 \to 2.1$ | $77.2 \pm 3.4$ | $82.0 \pm 5.4$ | $12.1 \pm 0.5$ | $16.9 \pm 1.9$ | [7, 8] |
| $2.1 \to 3.$ | $62.3 \pm 9.2$ | $53.6 \pm 4.9$ | $7.8 \pm 1.2$ | $6.7 \pm 0.6$ | [7, 8] |
| *Total Light I=1* | $454.0 \pm 10.6$ | $438.2 \pm 10.2$ | $172.8 \pm 2.2$ | $172.1 \pm 3.7$ | |
| *Average* | $446.1 \pm 10.4 \pm 7.9$ | | $172.5 \pm 3.0 \pm 0.3$ | | |
| **Light Isoscalar** | | | | | |
| *Below 1.93* | | | | | |
| $\omega$ | | $31.5 \pm 1.1$ | | $12.7 \pm 0.4$ | NWA [9] |
| $\phi$ | | $52.3 \pm 1.2$ | | $13.7 \pm 0.3$ | NWA [9] |
| $0.66 \to 1.93$ | | $11.6 \pm 3.0$ | | $2.7 \pm 0.7$ | $\sum$ exclusive [10] |
| *From 1.93 to 3* | | | | | |
| $\omega(1.42)$, $\omega(1.65)$ | | $9.4 \pm 1.4$ | | $1.2 \pm 0.2$ | BW [11, 9] |
| $\phi(1.68)$ | | $14.6 \pm 4.6$ | | $1.7 \pm .5$ | BW [11, 12, 9] |
| *Total Light I=0* | | $119.0 \pm 5.9$ | | $32.1 \pm 1.0$ | |
| **Heavy Isoscalar** | | | | | |
| $J/\psi(1S \to 4.415)$ | | $116.3 \pm 6.2$ | | $4.0 \pm 0.2$ | NWA [9] |
| $\Upsilon(1S \to 11.020)$ | | $12.7 \pm 0.5$ | | $0.1 \pm 0.0$ | NWA [9] |
| $T(349)$ | | $-(0.1 \pm 0.0)$ | | $\approx 0$ | NWA [9, 5] |
| *Total Heavy I=0* | | $128.9 \pm 6.2$ | | $4.1 \pm .2$ | |
| **QCD continuum** | | | | | |
| $3. \to 4.57^2$ | | $330.1 \pm 1.0$ | | $17.5 \pm .1$ | $(u,d,s)$ |
| $4.57^2 \to 11.27^2$ | | $503.0 \pm 1.0$ | | $5.0 \pm .1$ | $(u,d,s,c)$ |
| $11.27^2 \to (M_Z - 3 \text{ GeV})^2$ | | $2025.7 \pm 2.0$ | | $1.3 \pm 0.0$ | $(u,d,s,c,b)$ |
| $(M_Z + 3 \text{ GeV})^2 \to 4M_t^2$ | | $-(794.6 \pm 0.6)$ | | $\approx 0$ | $-$ |
| Z-pole | | $29.2 \pm .5$ | | $\approx 0$ | principal value [5] |
| $4M_t^2 \to \infty$ | | $-(24.0 \pm 0.1)$ | | $\approx 0$ | $(u,d,s,c,b,t)$ |
| *Total QCD Cont.* | | $2069.4 \pm 5.2$ | | $23.8 \pm 1.4$ | |
| **Final value** | **$2763.4 \pm 16.5$** | | **$232.5 \pm 3.2$** | | |

one obtains the final estimate:
$$\alpha^{-1}(M_Z) = 128.926(25) , \qquad (14)$$
which we show in Fig 1 for a comparison with recent existing determinations. One can notice an improved accuracy of the different recent determinations [3, 6, 5, 13] [2], which are in fair agreement with each others. Also a detailed comparison of each region of energy with the most recent work of [5] shows the same features (agreement and slight difference) like in the case of $a_\mu$ in SN, due to the slight difference in the parametrization of the data and spectral function. However, the final results are comparable. Finally, one can remark that due to the high-energy constant behaviour of the QED kernel function in this case, the low-energy region is no longer dominating. For $a_\mu$, the contribution of the $\rho$-meson below 1 GeV is 68% of the total contribution, while the sum of the QCD continuum is only 7.4% (see e.g. SN). Here the situation is almost reversed: the contribution of the $\rho$-meson below 1 GeV is only 2%, while the sum of the QCD-continuum is 73.6%. For this reason, improvement due to the new Novosibirsk $e^+e^-$ data [15] in the low-energy region will not be very significant. At present, new BES data [16] in the $J/\psi$ region are also available, which can be alternatively used. Below the $J/\psi$ resonances, the BES data are in excellent agreement with the QCD parametrization to order $\alpha_s^3$ used here for 3 flavours, justifying the accuracy of your input. Above the $J/\psi$ resonances, the parametrization used here (sum of narrow resonances +QCD continuum away from thresholds) can also be compared with these data. On can notice that, in the resonance regions, the BES data are more accurate than previous ones, which may indicate that our quoted errors in Table 1 for the $J/\psi$ family contributions are overestimated. In addition, the threshold of the QCD continuum which we have taken above the $J/\psi$ resonances, matches quite well with the one indicated by the BES data. We expect that with this new improved estimate of $\alpha(M_Z)$, strong constraint on the Higgs mass can be derived.

---
[2]Previous works are quoted in [14].

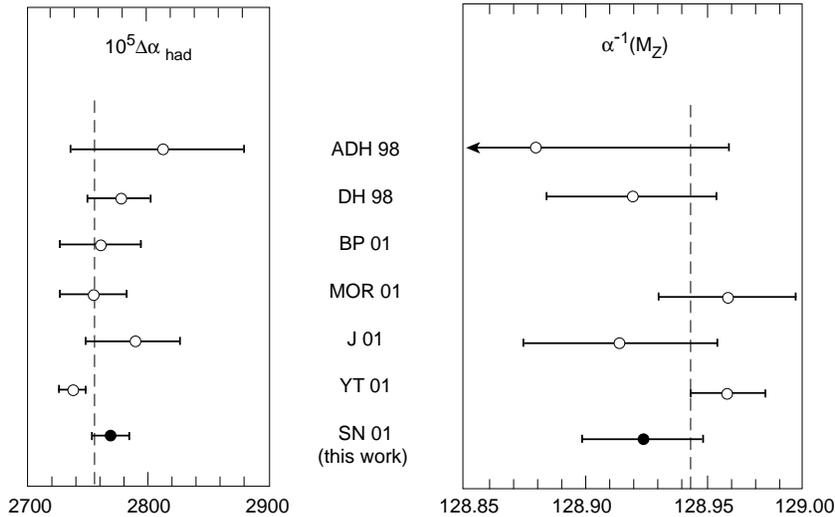

Figure 1: Recent determinations of $\Delta\alpha_{\text{had}}$ and $\alpha^{-1}(M_Z)$. The dashed vertical line is the mean central value. References to the authors are in [3, 5, 6, 13].

## 4 The Muonium Hyperfine Splitting

Our final result from Table 1:
$$\Delta\nu_{\text{had}} = (232.5 \pm 3.2) \text{ Hz} \qquad (15)$$
is shown in Table 2 in comparison with other determinations, where there is an excellent agreement with the most recent determination [4]. Here, due to the $(\ln t)/t$ behaviour of the kernel function, the contribution of the low-energy region is dominant. However, the $\rho$-meson region contribution below 1 GeV is 47% compared with the 68% in the case of $a_\mu$, while the QCD continuum is about 10% compared to 7.4% for $a_\mu$. The accuracy of our result is mainly due to the use of the $\tau$-decay data, explaining the similar accuracy of our final result with the one in [4] using new Novosibirsk data. The agreement with [4] canbe understood from the agreement of the averaged correlated $e^+e^-$ and $\tau$-decay data compiled in [6] with the new Novosibirsk data used in [4]. We differ with DH98 [13] in the treatment of the QCD contribution [3]. For light quarks, QCD is applied in the region where non-perturbative contributions are inessential. For heavy quarks, perturbative QCD is applied far from heavy quark thresholds, where it can be unambiguously used. Adding to this result, the QED contribution up to fourth order, the lowest order electroweak contribution [4], and an estimate of the higher order weak and hadronic contributions [18]:

$$\Delta\nu_{\text{QED}} = 4\,270\,819(220) \text{ Hz}$$
$$\Delta\nu_{\text{weak}}(l.o) = -\frac{G_F}{\sqrt{2}} m_e m_\mu \left(\frac{3}{4\pi\alpha}\right) \nu_F \simeq -65 \text{ Hz} ,$$

Table 2: Recent determinations of $\Delta\nu_{\text{had}}$

| Authors | $\Delta\nu_{\text{had}}$ [Hz] |
|---|---|
| FKM 99 [17] | $240 \pm 7$ |
| CEK 01 [4] | $233 \pm 3$ |
| SN 01 (This work) | $232.5 \pm 3.2$ |

---
[3]For more details, see [1].

$$\begin{aligned} |\Delta\nu_{\text{weak}}(h.o)| &\approx 0.7 \text{ Hz}, \\ \Delta\nu_{\text{had}}(h.o) &\simeq 7(2) \text{ Hz}, \end{aligned} \qquad (16)$$

one obtains the Standard Model (SM) prediction:

$$\nu_{\text{SM}} \equiv \nu_F + \Delta\nu_{\text{QED}} + \Delta\nu_{\text{weak}} + \Delta\nu_{\text{had}} + \Delta\nu_{\text{had}}(h.o). \qquad (17)$$

If one uses the relation:

$$\nu_F = \rho_F \left(\frac{\mu_\mu}{\mu_B^e}\right) \frac{1}{(1+m_e/m_\mu)^3}, \qquad (18)$$

with:

$$\rho_F = \frac{16}{3}(Z\alpha)^2 Z^2 cR_\infty, \qquad (19)$$

one would obtain:

$$\nu_{\text{SM}} = 4\,463\,302\,913(511)(34)(220) \text{ Hz}, \qquad (20)$$

where the two first errors are due to the one of the Fermi splitting energy. The first largest one being induced by the one of the ratio of the magnetic moments. The third error is due to the 4th order QED contribution where, one should notice that, unlike the case of $a_\mu$, the dominant errors come from the QED calculation which should then be improved. Unfortunately, these previous errors are still too large and obscure the effects of the electroweak and hadronic contributions. One the opposite, the data are very precise [19]:

$$\nu_{\text{exp}} = 4\,463\,302\,776(51) \text{ Hz}. \qquad (21)$$

Therefore, at present, we find, it is more informative to extract the Fermi splitting energy $\nu_F$ from a comparison of the Standard Model (SM) prediction with the experimental value of $\nu$. Noting that $\nu_F$ enters as an overall factor in the theoretical contributions, one can rescale the previous values and predict the ratio:

$$\frac{\nu_{\text{SM}}}{\nu_F} = 1.000\,957\,83(5). \qquad (22)$$

Combining this result with the previous experimental value of $\nu$, one can deduce the SM prediction:

$$\nu_F = 4\,459\,031\,783(226) \text{ Hz}, \qquad (23)$$

where the error is dominated here by the QED contribution at fourth order. However, this result is a factor two more precise than the determination in [4] given in Eq. (8), where the main error in Eq. (8) comes from the input values of the magnetic moment ratios. Using this result in Eq. (23) into the expression:

$$\nu_F = \rho_F \left(\frac{m_e}{m_\mu}\right) \frac{1}{(1+m_e/m_\mu)^3} (1+a_\mu), \qquad (24)$$

where:

$$\rho_F = \frac{16}{3}(Z\alpha)^2 Z^2 cR_\infty, \qquad (25)$$

and $Z=1$ for muonium, $\alpha^{-1}(0)=137.035\,999\,58(52)$ [20], $cR_\infty = 3\,289\,841\,960\,368(25)$ kHz [21] and $a_\mu = 1.165\,920\,3(15) \times 10^{-3}$ [22], one can extract a value of the ratio of the muon over the electron mass:

$$\frac{m_\mu}{m_e} = 206.768\,276(11), \qquad (26)$$

to be compared with the PDG value 206.768 266(13) using the masses in MeV units, and with the one from [4]: 206.768 276(24). After inserting the previous value of $m_e/m_\mu$ into the alternative (equivalent) relation:

$$\nu_F = \rho_F \left(\frac{\mu_\mu}{\mu_B^e}\right) \frac{1}{(1+m_e/m_\mu)^3}, \qquad (27)$$

one can deduce the ratio of magnetic moments:

$$\frac{\mu_\mu}{\mu_B^e} = 4.841\,970\,47(25) \times 10^{-3}, \qquad (28)$$

compared to the one obtained from the PDG values of $\mu_\mu/\mu_p$ and $\mu_p/\mu_B^e$ [9]: $\mu_\mu/\mu_B^e = 4.841\,970\,87(14) \times 10^{-3}$. In both applications, the results in Eqs. (26) and (28) are in excellent agreement with the PDG values. These remarkable agreements can give strong constraints to some contributions beyond the Standard Model and should be explored.

## 5  Conclusions

We have evaluated the hadronic and QCD contributions $\Delta\alpha_{\text{had}}$ and $\Delta\nu_{\text{had}}$ respectively to the QED running coupling and to the Muonium hyperfine splitting. Our results shown in Eqs. (11) and (15), are in excellent agreement with existing determinations shown in Fig. 1 and Table 2 and are quite accurate. These results have been obtained within the same strategy and data input as the one of the anomalous magnetic moment obtained previously in SN [1]. For this reason, they are self-contained outputs. One of the immediate consequences of these results is the prediction of $\alpha(M_Z)$ given in Eq. (14), while we have used the result for the muonium hyperfine splitting for a high precision measurement of the ratios of the muon over the electron mass given in Eq. (26) and of magnetic moments given in Eq. (28). These Standard Model predictions are in excellent agreement with the ones quoted by PDG [9]. These agreements can be used for providing strong constraints on some model buildings beyond the Standard Model.

## Acknowledgements

It is a pleasure to thank W-Y. Pauchy Hwang for the hospitality at CosPA-NTU (Taipei), where this work has been done.